\def\Date{December 21, 2017}  

\documentclass[a4paper,11pt]{article}

\newcount\relabeling\relabeling=1  
\usepackage[T1]{fontenc}
\usepackage[latin2]{inputenc}
\usepackage{amsmath,amsfonts,amssymb}
\newcommand\nc{\newcommand*}  \nc\ncc{\newcommand}  \nc\rc{\renewcommand}

\ncc\OMIT[1]{\relax}  
\ncc\VOMIT[1]{#1}     

\DeclareMathAlphabet{\mathpzc}           
  {OT1}{pzc}{m}{it}  \nc\Cal\mathpzc    

\ncc\quot[1]{`#1'}  \ncc\quott[1]{``#1''}

\nc\eq[2]{\begin{align} \label{#1} #2 \end{align}}
\nc\lel[1]{\\ \label{#1}}
\nc\non{\tag*{}}
\nc\re[1]{(\ref{#1})}
\nc\m[1]{$            #1         $}  
\nc\mm[1]{$   \,      #1  \,     $}  
\nc\mmm[1]{$  \,\,    #1  \,\,   $}  
\nc\mmmm[1]{$ \,\,\,  #1  \,\,\, $}  
\nc\f\frac
\nc\ff[2]{{\textstyle \f{#1}{#2}}}
\nc\mat[1]{\begin{matrix} #1 \end{matrix}}  
\nc\smat[1]{\begin{smallmatrix} #1 \end{smallmatrix}}  
\nc\mS[1]{\ifcase#1\displaystyle\or\textstyle  
  \or\scriptstyle\or\scriptscriptstyle\else\textstyle\fi}

\nc\0[2]{\ifcase#1{#2}\or\lt(#2\rt)\or\lt[{#2}\rt]\or\lt\{{#2}\rt\}\or
  \mathord<{#2}\mathord>\or\lt\langle{#2}\rt\rangle\or\lt\lvert{#2}\rt
  \rvert\or\lt\lVert{#2}\rt\rVert\fi}
\nc\1[2]{\ifcase#1{#2}\or(#2)\or[#2]\or\{#2\}\or\mathord<{#2}\mathord
  >\or\langle{#2}\rangle\or\lvert{#2}\rvert\or\lVert{#2}\rVert\fi}
\nc\2[2]{\ifcase#1{#2}\or\big(#2\big)\or\big[#2\big]\or\big
  \{#2\big\}\or\big<#2\big>\or\big\langle#2\big\rangle\or\big
  \lvert#2\big\rvert\or\big\lVert#2\big\rVert\fi}
\nc\3[2]{\ifcase#1{#2}\or\Big(#2\Big)\or\Big[#2\Big]\or\Big\{#2\Big
  \}\or\Big<#2\Big>\or\Big\langle#2\Big\rangle\or\Big\lvert#2\Big
  \rvert\or\Big\lVert#2\Big\rVert\fi}
\nc\4[2]{\ifcase#1{#2}\or\bigg(#2\bigg)\or\bigg[#2\bigg]\or\bigg
  \{#2\bigg\}\or\bigg<#2\bigg>\or\bigg\langle#2\bigg\rangle\or\bigg
  \lvert#2\bigg\rvert\or\bigg\lVert#2\bigg\rVert\fi}
\nc\5[2]{\ifcase#1{#2}\or\Bigg(#2\Bigg)\or\Bigg[#2\Bigg]\or\Bigg
  \{#2\Bigg\}\or\Bigg<#2\Bigg>\or\Bigg\langle#2\Bigg\rangle\or\Bigg
  \lvert#2\Bigg\rvert\or\Bigg\lVert#2\Bigg\rVert\fi}
\nc\9[2]{\ifcase#1{#2}\or\left(#2\right)\or\left[#2\right]\or\left
  \{#2\right\}\or\left\langle{#2}\right\rangle\or\left\langle{#2}\right
\rangle\or\left\lvert{#2}\right\rvert\or\left\lVert{#2}\right\rVert\fi}
\nc\lt{\mathopen{}\mathclose\bgroup\left}
\nc\rt{\aftergroup\egroup\right}

\nc\Mathbf[1]{\mathchoice   
  {\hbox{\boldmath{$#1$}}}  {\hbox{\boldmath{$#1$}}}
  {\hbox{\boldmath{$\scriptstyle #1$}}}
  {\hbox{\boldmath{$\scriptscriptstyle #1$}}}}

\nc\x\hskip       \nc\y\vskip       \nc\X\kern       
\nc\xph\hphantom  \nc\yph\vphantom  \nc\ph\phantom
\nc\XX[1]{\HB to#1{\ }}  \nc\YY[1]{\setbox1\HB to0em{\ }\RB{#1}}
\nc\bi\relax                      
\nc\bit{      \mskip1mu}  \nc\biT{      \mskip-1mu}  
\nc\bitt{     \mskip2mu}  \nc\biTT{     \mskip-2mu}  
\nc\bittt{    \mskip3mu}  \nc\biTTT{    \mskip-3mu}  
\nc\bitttt{   \mskip4mu}  \nc\biTTTT{   \mskip-4mu}  
\nc\bittttt{  \mskip5mu}  \nc\biTTTTT{  \mskip-5mu}  
\nc\bitttttt{ \mskip6mu}  \nc\biTTTTTT{ \mskip-6mu}
\nc\bittttttt{\mskip7mu}  \nc\biTTTTTTT{\mskip-7mu}
\nc\HB\hbox  \nc\SB{\setbox1\HB}  \nc\CB{\copy1}  
             \nc\SC{\setbox2\HB}  \nc\CC{\copy2}  
\nc\RB[1]{\raise#1\CB}  \nc\XB{\wd1}  \nc\YB{\ht1}  \nc\ZB{\dp1}
\nc\RC[1]{\raise#1\CC}  \nc\XC{\wd2}  \nc\YC{\ht2}  \nc\ZC{\dp2}
\nc\UB{\Z-\XB}  \nc\VB{\CB\UB}  \nc\WB[1]{\RB{#1}\UB}  
\nc\UC{\Z-\XC}  \nc\VC{\CC\UC}  \nc\WC[1]{\RC{#1}\UC}  

\newdimen\w
\ncc\Parbox[1]{\parbox[t]{\w}{\baselineskip3. ex #1}}
\nc\qheaders[1]{\textbf{#1}}
\nc\qhj{c}
\nc\qheader[2]
  {\multicolumn{1}{c}{\underline{\makebox[\w][\qhj]{\qheaders{#1}}}}
  &\multicolumn{1}{c}{\underline{\makebox[\w][\qhj]{\qheaders{#2}}}}\\}
\nc\Yy{\rule{0em}{3.5 ex}&\\}
\ncc\qrow[2]{\Yy \Parbox{#1}&\Parbox{#2}\\}

\nc\tensUpRt[1]{^{\mathrm{#1}}}            
\nc\symm{\tensUpRt{S}}   \nc\asymm{\tensUpRt{A}}

\nc\tensor\mathbf  
\nc\Tensor\Mathbf  
\nc\four[1]{\hat{\tensor{#1}}}

\nc\qmat[1]{\begin{pmatrix}#1\end{pmatrix}}
\nc{\Tr}{\mathop{\mathrm{Tr}}\nolimits}
\nc\scal[2]{\1 5 {#1, #2}}  \nc\Scal[2]{\9 5 {#1, #2}}
\makeatletter
\nc\sle{\mathord{\newmcodes@\kern\z@\textsl e\x.06em}}  
\nc\sli{\mathord{\newmcodes@\kern\z@\textsl i\x.06em}}
\makeatother
\nc\e\sle  \nc\ii\sli   
\nc\dd{\mathrm{d}} \nc\ddd{\bit\d} 
\nc\pd\partial
\nc\qdag{\dag}          \nc\qdagq{\mathrm{?}}
\nc\qtrans{\mathrm{T}}
\nc\qbar{\overline}
\nc\intt[1]{\int}
\nc\qnull{0}  \nc\qone{I_2}  \nc\qqone{I_4}
\nc\qqnull{\mathbf{0}}
\nc\qrepl{\leadsto}

\nc\qat{\bittttttt}
\nc\qnl{\rule{0em}{2.4 ex} \\}

\nc\qA{A}
\nc\qC{\mathbb{C}}
\nc\qD{D}
\nc\qP{\Cal{P}}
\nc\qPp{\qP}
\nc\qH{\Cal{H}}
\nc\qHfpp{\qH^{4,+}_{\qPp}}  \nc\qHfp{\qH^4_{\qPp}}
\nc\qHfxp{\qH^{4,+}_{\qX}}   \nc\qHfx{\qH^4_{\qX}}
\nc\qHffpp{\qH^{4 \wedge 4,+}_{\qPp}}  \nc\qHffp{\qH^{4 \wedge 4}_{\qPp}}
\nc\qHffxp{\qH^{4 \wedge 4,+}_{\qX}}   \nc\qHffx{\qH^{4 \wedge 4}_{\qX}}
\nc\qL{L}                 \nc\qqL{\Cal{L}}
\nc\qNpp{N_+ \0 1 {\qp}}  \nc\qNmp{N_- \0 1 {\qp}}  \nc\qNpmp{N_\pm \0 1 {\qp}}
\nc\qNppn{N_+ \0 1 {\qpn}}  \nc\qNmpn{N_-\0 1 {\qpn}}
\nc\qU{U}
\nc\qW{W}
\nc\qX{\Cal{X}}

\nc\qa{a}  \nc\qb{b}  
\nc\qqa{a}
\nc\qc{c}
\nc\qe{e}
\nc\qg{g}
\nc\qj{j}  \nc\qk{k}  
\nc\qqj{j}            
\nc\qm{m}
\nc\qn{n}
\nc\qp{p}  \nc\qqp{\mathbf{\qp}}
\nc\qpn{{\check\qp}}
\nc\qt{t}
\nc\qx{x}  \nc\qqx{\mathbf{\qx}}

\nc\qxi{\theta}  \nc\qXi{\Theta}
\nc\qPhi{\qzetx}
\nc\qalp{\alpha}  \nc\qbet{\beta}  \nc\qgam{\gamma}
\nc\qmu{\mu}  \nc\qnu{\nu}  
\nc\qchi{\chi}
\nc\qsig{\sigma}
\nc\qphip{\phi}     \nc\qphix{\phi}
\nc\qpsip{\psi}     \nc\qpsix{\psi}
\nc\qzetp{\zeta}    \nc\qzetx{\zeta}

\begin{document}

\nc\qnls{\\ \footnotesize}
\title{Probability current in zero-spin relativistic quantum mechanics}
\author{%
Tam\'as F\"ul\"op \qnls
Department of Energy Engineering \qnls
Budapest University of Technology and Economics \qnls
3 M\H uegyetem rkp., 1111 Budapest, Hungary \qnls
fulop@energia.bme.hu
 \\ \\
Tam\'as Matolcsi \qnls
Department of Applied Analysis and Computational Mathematics \qnls
E\"otv\"os Lor\'and University \qnls
1/C P\'azm\'any P. Stny., 1117 Budapest, Hungary
}

\date{\Date}

\maketitle

 \abstract{%
We show that the antisymmetric spinor tensor representation of spin-0
relativistic quantum mechanics provides a conserved current with
positive definite timelike component,
interpretable as probability density.
The construction runs in complete analogy to
the spin-1/2 case, and
provides an
analogously
natural one-particle Hilbert space description for
spin 0.
Except for the free particle, the obtained formulation
proves to be inequivalent to the
one based on the Klein--Gordon equation.
%
The second quantized version may
 lead to
new field theoretical interaction terms for zero-spin particles.%
\footnote{Material presented at the  
Zimányi School, December 7-11, 2015, Budapest, Hungary.%
}

\section{Introduction}\label{.1..1.}

The Klein--Gordon equation for a scalar wave function,
 \eq{.1.1.}{
\9 2 { \9 1 { \pd^\qmu + \ii \qe \qA^\qmu } \9 1 { \pd_\qmu + \ii
\qe \qA_\qmu } - \qm^2 } \qphix = 0
 }
\1 1 {see notation conventions below} is very plausible relativistic
quantum mechanical model for a spin-0 particle in an external
four-potential field but the corresponding conserved four current
 \eq{.1.2.}{
\f {1}{2 \ii \qm} \9 1 { \qphix^* \pd^\mu \qphix - \pd^\mu \qphix^*
\bit \qphix }
 }
cannot be interpreted as a probability current since its timelike
component is not positive definite. While the so-called
Feshbach--Villars formalism \cite{FesVil} rewrites \re{.1.1.} as a
Schrödinger-type -- first-order in time derivative -- equation on the
two-component wave function
 \eq{.1.3.}{
\qmat{ \qphix + \f {\ii}{\qm} \9 1 { \pd_0 + \ii e A_0 } \qphix \\
\qphix - \f {\ii}{\qm} \9 1 { \pd_0 + \ii e A_0 } \qphix
\rule{0em}{3.4ex} } ,
 }
and can restrict solutions to those with positive integral of the
timelike component of \re{.1.2.}, a Hilbert space structure cannot be
established.

For the free particle, the \quot{positive} solutions of
 \eq{.1.4.}{
\9 1 { \pd^\qmu \pd_\qmu - \qm^2 } \qphix = 0
 }
satisfy \cite{Jost}
 \eq{.1.5.}{
\ii\bit \pd_0 \qphix = \sqrt{- \triangle + m^2} \bittt \qphix
 }
and admit a time independent Hilbert space scalar product
 \eq{.2.6.}{
\Scal{\qphix_1
}{\qphix_2
} = \intt{\qX} \qphix_1^\qdag \qphix_2^{} \bittt \dd^3 \qqx
 }
but its generalization
 \eq{.2.7.}{
\ii\bi \9 1 { \pd_0 + \ii e A_0 } \qphix = \sqrt{ \9 1 { \pd_\qj + \ii e
A_\qj } \9 1 { \pd_\qj + \ii e A_\qj } + m^2} \bittt \qphix
 }
to \m { \qA_\qmu \ne 0 } is not equivalent to \re{.1.1.},
 \eq{.2.8.}{
\9 2 {  \9 1 { \pd^\qmu + \ii \qe \qA^\qmu } \9 1 { \pd_\qmu + \ii
\qe \qA_\qmu } - \qm^2 } \qphix \ne 0
 }
since \m { \pd_\mu } and \m { A_\nu } do not commute.

The situation is in sharp contrast with the spin-1/2 case where the
Dirac equation automatically provides \quot{positive} solutions, a
conserved current with positive definite timelike component, and a
Hilbert space structure, all for \m { \qA_\qmu \ne 0 } as well.

In this writing, we show that the antisymmetric spinor tensor
representation of free spin-0 particles \1 1 {see, e.g., \cite{Var}} can
be generalized to \m { \qA_\qmu \ne 0 } with a conserved current with
positive definite timelike component, and a corresponding Hilbert
space. In \cite{Var}, only the free system is presented and only in
momentum space -- here, we perform the transformation to coordinate
space, and carry out the generalization \m { \qA_\qmu \ne 0 }.

The construction can be established in complete analogy to the spin-1/2
case, and provides an analogously natural consistent one-particle theory
for spin 0.

The free particle case is shown, via the adaptation of the
spin-1/2 Foldy--Wouthuysen transformation, to be equivalent to the
\re{.1.5.} version of scalar Klein--Gordon quantum mechanics, while
for nonzero external field the equivalence is broken. Accordingly, we
expect that, e.g., the Coulomb problem admits a spectrum different from
the scalar Klein--Gordon one.

We start with revisiting the case of spin 1/2. This review is
intentionally detailed and pedagogical -- the spin-0 version can then be
presented in a straightforward step-by-step way,
due
to the strong analogy between the two situations.

\section{Basics}\label{.2..2.}

\subsection{Notations}\label{.2..2.1.}

In most respects, our notations follow those of \cite{Greiner}. We
work in the convention \m { \hbar \equiv c \equiv 1 }, consider the
Lorentz metric \m { \qg } with signature \m { \1 1 { \mathrel + \bit
\mathrel - \bit \mathrel - \bit \mathrel - }
}, and use spacetime four-indices \m { \qmu, \qnu = 0, 1, 2, 3 } 
and
three-indices \m { \qj, \qk = 1, 2, 3 }%
. Repeated indices involve summation. Complex conjugate is denoted by \m
{ ^* \bit}, and its combination with transposition \m { ^\qtrans } is
indicated by \m { ^\qdag }.
With the Pauli matrices and the \m { 2 \times 2 } unit matrix,
 \eq{.2.9.}{
\qsig_1 = \qmat{ 0 & 1 \\ 1 & 0 } ,
 \qquad
\qsig_2 = \qmat{ 0 & -\ii \\ \ii & 0 } ,
 \qquad
\qsig_3 = \qmat{ 1 & 0 \\ 0 & -1 } ,
 \qquad
\qone = \qmat{ 1 & 0 \\ 0 & 1 } ,
 }
we introduce the \m { 4 \times 4 } matrices
 \eq{.2.10.}{
\qbet = \qgam_0 = \qmat{ \qone & \qnull \\ \qnull & - \qone } ,
 \qquad
\qgam_\qj = \qmat{ \qnull & - \qsig_\qj \\ \qsig_\qj & \qnull } ,
 \qquad
\qalp_\qj = - \qbet \qgam_\qj =
\qmat{ \qnull & \qsig_\qj \\ \qsig_\qj & \qnull } ,
 }
where \m { \qnull } denotes the \m { 2 \times 2 } zero matrix as well.
These possess the properties
 \eq{.2.11.}{
\qgam_\qmu \qgam_\qnu + \qgam_\qnu\qgam_\qmu = 2 \qg_{\qmu\qnu} \qqone ,
 \qquad
\qgam_0^\qdag = \qgam_0^{} ,
 \qquad
\qgam_\qj^\qdag = - \qgam_\qj^{} ,
 \qquad
\qgam_\qmu^\qdag \qbet = \qbet \qgam_\qmu^{} ,
 \qquad
\qalp_\qj^\qdag = \qalp_\qj^{} .
 }
For elements \m { \qxi } of \m { \qC^4 }, as well as for complex
\m { 4 \times 4 } matrices \m { \qXi }, we put
 \eq{.2.12.}{
\qbar\qxi = \qxi^\qdag \qbet ,
 \qquad
\qbar\qXi = \qXi^\qdag \qbet .
 }

We discuss the case of positive particle mass, \m { \qm > 0 } only.
Four-momenta \m { \qp } with
 \eq{.2.13.}{
\qp_0 = \sqrt{ \qp_\qj \qp_\qj + \qm^2 } = \sqrt{ \qqp^2 + \qm^2 }
 }
form the set \m { \qPp } -- the positive mass shell --, on which the
Lorentz invariant integration measure
is \m { \f {\qm}{\qp_0} \dd^3 \qqp }
 (up to proportionality).

\subsection{Geometric ingredients}\label{.3..2.2.}

For any \m { \qp } from \m { \qPp }, the eigenvalues of the matrix
 \eq{.3.14.}{
\qp_\qmu \qgam^\qmu
 }
are \m { \qm } and \m { -\qm }, following from that \mm { \9 1 {
\qp_\qmu \qgam^\qmu }^2 = \qm^2 \qqone }; and the corresponding two
eigensubspaces \m { \qNpp }, \m { \qNmp } are both two-dimensional. For
example, especially simple is the case  of the four-momentum that is at
rest with respect to the inertial reference frame used:
 \eq{.3.15.}{
\qpn = \qmat{ \qm \\ \qqnull } ,
 \qquad
\qpn_\qmu \qgam^\qmu = \qm \qgam_0 = \qm \qbet :
 }
then
 \eq{.3.16.}{
\qNppn \text{ is spanned by}
 \quad
\qmat{ 1 \\ 0 \\ 0 \\ 0 } ,  \quad  \qmat{ 0 \\ 1 \\ 0 \\ 0 } ,
 \qquad
 \qquad
\qNmpn \text{ is spanned by}
 \quad
\qmat{ 0 \\ 0 \\ 1 \\ 0 } ,  \quad  \qmat{ 0 \\ 0 \\ 0 \\ 1 } .
 }
For other \m { \qp }s, upper and lower components become mixed.

One finds, analogously, that the same \m { \qNpp } and \m { \qNmp } are
the eigensubspaces of
 \eq{.3.17.}{
\qalp_\qj \qp_\qj + \qbet \qm ,
 }
with eigenvalues \m { \qp_0 } and \m { - \qp_0 }, respectively.

\section{Spin 1/2 quantum mechanics}\label{.3..3.}

\subsection{Free particle, momentum space}\label{.3..3.1.}

The momentum space version of the Dirac equation,
 \eq{.3.18.}{
\qgam^\qmu\qp_\qmu \qpsip \1 1 { \qp } = \qm \qpsip \1 1 { \qp }
 }
has, in the light of the previous section, the simple geometric
interpretation that, at each \m { \qp }, \m { \qpsip \1 1 { \qp } } has
to be an element of \m { \qNpp },
 \eq{.3.19.}{
\qgam^\qmu \qp_\qmu \qpsip \1 1 { \qp } = \qm \qpsip \1 1 { \qp }
\quad \Longleftrightarrow \quad
 \qpsip \1 1 { \qp } \in \qNpp .
 }
For example, at  \m { \qpn = \qmat{ \qm \\ \qqnull } }, a solution \m {
\qpsip \0 1 { \qpn } } can have only upper nonzero components,
 \eq{.3.20.}{
\qpsip \2 1 { \qpn } = \qmat{ \qpsip_1 \2 1 { \qpn }  \qnl
\qpsip_2 \2 1 { \qpn }  \qnl  0  \qnl  0 } .
 }
For other \m { \qp }s, upper and lower components become mixed but there
are still only two degrees of freedom -- when we expand \m { \qpsip \1 1
{\qp} } with respect to basis vectors \m { \qn_1,
\qn_2, \qn_3, \qn_4 } in \m { \qC^4 } where
\m { \qn_1 } and \m { \qn_2 } are in \m { \qNpp } and \m { \qn_3, \qn_4 }
are in \m { \qNmp },
 \eq{.3.21.}{
\qpsip \1 1 {\qp} = \qc_1 \qn_1 + \qc_2 \qn_2 + \qc_3 \qn_3 + \qc_4 \qn_4 ,
 \qquad
\qn_1, \qn_2 \in \qNpp , \quad \qn_3, \qn_4 \in \qNmp ,
 }
then
 \eq{.3.22.}{
\9 1 { \qgam^\qmu \qp_\qmu - \qm } \qpsip\1 1 {\qp} = \9 1 { \qgam^\qmu
\qp_\qmu - \qm } \9 1 { \qc_1 \qn_1 +
\cdots + \qc_4 \qn_4 } = 0
 \quad \Longrightarrow \quad
\qc_3 = \qc_4 = 0 ,
 }
and only the two coefficients \m { \qc_1 }, \m { \qc_2 } can be nonzero.

Being on the positive mass shell \re{.2.13.} involves
 \eq{.3.23.}{
\qp_0 \qpsip = \sqrt{ \qp_\qj \qp_\qj + \qm^2 } \bittt \qpsip
 }
from which
 \eq{.3.24.}{
\qp_0^2 \qpsip = \9 1 { \qp_\qj \qp_\qj + \qm^2 } \qpsip
 }
follows so we obtain the momentum space version of the Klein--Gordon
equation,
 \eq{.4.25.}{
\9 1 { \qp^\qmu \qp_\qmu - \qm^2 } \qpsip = 0 .
 }
The same conclusion can be derived via acting by \mm { \qgam^\qmu
\qp_\qmu + \qm } on \mm { \9 1 { \qgam^\qmu\qp_\qmu \qpsip - \qm }
\qpsip = 0 } \1 2 {see \re{.3.18.}} from the left.

To each proper Lorentz transformation \m { \qL } there exists \1 1 {see,
\textit{e.g.,} \cite{MTnarancs}} -- uniquely up to a unit multiplier --
a \m { 4 \times 4 } matrix \m { \qD_\qL } such that
 \eq{.4.26.}{
\qbet \qD_\qL^\qdag \qbet = \qD_\qL^{-1} ,
 \qquad
\qD_\qL^{} \9 1 { \qgam^\qmu \qp_\qmu } \qD_\qL^{-1} = \qgam^\qmu \1 1 {
\qL \qp }_\qmu .
 }
As a consequence, \m { \qbar{\qD_\qL \qxi} \bit \qD_\qL \qxi' =
\qbar{\qxi} \bit \qxi' }. Proper Lorentz transformations map
four-momenta of \m { \qPp } to four-momenta still within \m { \qPp },
and the formula
 \eq{.4.27.}{
\9 1 { \qU_{\1 1 { \qa, \qL }} \qpsip } \1 1 { \qp } =
\e^{ \ii \qp_\qmu \qa^\qmu } \qD_\qL \qpsip \1 1 { \qL^{-1} \qp }
 }
proves \cite{Var,MTnarancs} to give the spin-1/2 irreducible unitary ray
representation of the proper Poincar\'e group (\m { \qa }: a
translation, \m { \qL }: a Lorentz transformation) on the Hilbert space
\m { \qHfpp } of (measurable) \m { \qC^4 }  valued functions \m { \qpsip
} defined on \m { \qPp } with \m { \qpsip \1 1 { \qp } } being in \m {
\qNpp } for each \m { \qp }, and with finite integral
 \eq{.4.28.}{
\intt {\qPp} \qbar{\qpsip\1 1 { \qp }} \qpsip\1 1 { \qp }
\f {\qm \bit \dd^3 \qqp}{\qp_0} = \intt {\qPp} {\qpsip\1 1 { \qp }}^\qdag
\qpsip\1 1 { \qp } \f {\qm^2 \bi \dd^3 \qqp}{\qp_0^2} .
 }
Why these two integrals are equal follows from that, for \m { \qpsip \1
1 { \qp } } in \m { \qNpp },
 \eq{.4.29.}{
\9 1 { \qalp_\qj \qp_\qj + \qbet \qm } \qpsip \1 1 { \qp } & = \qp_0
\qpsip \1 1 { \qp } ,
 }
 \eq{.4.30.}{
\qsig_\qj \qp_\qj \qmat{ \qpsip _1 \\ \qpsip _2 } = \9 1 { \qp_0 + \qm }
\qmat{ \qpsip _3 \\ \qpsip _4 } ,
 \qquad
\qsig_\qj \qp_\qj \qmat{ \qpsip _3 \\ \qpsip _4 } = \9 1 { \qp_0 - \qm }
\qmat{ \qpsip _1 \\ \qpsip _2 } ,
 }
 \eq{.4.31.}{
\qbar{\qpsip} \qpsip = \f {2\qm}{\qp_0 + \qm} \9 1 { \9 6 { \qpsip_1 }^2
+ \9 6 { \qpsip_2 }^2 } ,
 \qquad
\qpsip^\qdag \qpsip = \f {2\qp_0}{\qp_0 + \qm} \9 1 { \9 6 { \qpsip_1 }^2
+ \9 6 { \qpsip_2 }^2 } ,
 }
 \eq{.4.32.}{
\qbar{\qpsip\1 1 { \qp }} \qpsip\1 1 { \qp } = \f {m}{\qp_0} {\qpsip\1 1
{ \qp }}^\qdag \qpsip\1 1 { \qp } .
 }
The first form of the integral -- the lhs of \re{.4.28.} --  makes
Lorentz invariance apparent while the second \1 2 {the rhs of
\re{.4.28.}} emphasizes positive definiteness of the integrand. It is
important to bear in mind that, in the integrals, \m { \qpsip } is not
an arbitrary square integrable \m { \qC^4 } valued function -- that
would mean a larger Hilbert space \m { \qHfp } -- but has to be in the
subspace \m { \qHfpp } of that larger Hilbert space \m { \qHfp },
defined by the condition \m { \qpsip \1 1 { \qp } \in \qNpp }. It is the
solution space of \re{.3.19.} that becomes a Hilbert space by the
integral \re{.4.28.}.

In a rephrased form,  the multiplier operator \m { \qalp_\qj \qp_\qj +
\qbet \qm } is self-adjoint in \m{ \qHfp }, and has the spectrum \m { (
-\infty, -\qm ] \cup [ \qm, \infty ) }. The proper Hilbert space \m {
\qHfpp } is the subspace corresponding to the positive half of this
spectrum.

The physical meaning of being a representation of the Poincaré group is
to be a free system on special relativistic spacetime \1 1 {not to be
connected to anything distinguished}. The physical
meaning of being an irreducible representation is that the system is
elementary, not some composite one \1 1 {not some decomposable one}.

Except for \m { \qp = \qpn } \1 2 {recall \re{.3.20.}}, elements of \m {
\qNpp } have some nonzero third and/or fourth component. If \m { \qL_\qp
} denotes the Lorentz boost that brings \m { \qp } to \m { \qpn } then
\m { \qW \1 1 { \qp } \equiv \qD_{\qL_\qp} }, the so-called
Foldy--Wouthuysen transformation \cite{Pryce,Tani,Tanii,FW}, transforms
elements of \m { \qNpp } to elements of \m { \qNppn }, in other words, to
upper components only:
 \eq{.4.33.}{
\qpsip = \qmat{ \qpsip_1 \qnl \qpsip_2 \qnl \qpsip_3 \qnl \qpsip_4 } ,
 \qquad
\qpsip^{\qW} \1 1 { \qp } = \qW \1 1 { \qp } \bit \qpsip \1 1 { \qp } ,
 \qquad
\qpsip^{\qW} = \qmat{ \qpsip_1^{\qW} \qnl \qpsip_2^{\qW} \qnl 0 \qnl 0 } .
 }
As \m { \qD_{\qL} } is unique only up to unitary equivalence, so is \m
{\qW \1 1 { \qp } }. One possible choice is
 \eq{.4.34.}{
\qW \1 1 { \qp } = \f{1}{ \sqrt{ 2 \qm \9 1 {\qp_0 + \qm } } }
\2 2 { \1 1 { \qp_0 + \qm } \qqone - \qalp_\qj \qp_\qj } .
 }

Since \mm { \9 1 { \qp^\qmu \qp_\qmu - \qm^2 } \qqone } commutes with \m
{ \qW \1 1 { \qp } }, from \re{.4.25.} we find that
 \eq{.5.35.}{
\9 1 { \qp^\qmu \qp_\qmu - \qm^2 } \qpsip^{\qW} = 0
 }
also holds. In parallel, utilizing \re{.4.33.}, \re{.4.28.} can be
further expressed as
 \eq{.5.36.}{
\intt {\qPp} \qbar{\qpsip } \bitt \qpsip \bittt \f {\qm \bit \dd^3
\qqp}{\qp_0} = \intt {\qPp} {\qpsip }^\qdag \bi \qpsip \bittt \f {\qm^2
\bi \dd^3 \qqp}{\qp_0^2} = \intt {\qPp} \9 1 { \9 6 { \qpsip_1^{\qW} }^2
+ \9 6 { \qpsip_2^{\qW} }^2 } \f {\qm^2 \bi \dd^3 \qqp}{\qp_0^2} .
 }

\subsection{Free particle, coordinate space}\label{.5..3.2.}

Fourier transformation transports elements of \m { \qHfpp } to
coordinate space:
 \eq{.5.37.}{
\f{1}{(2\pi)^2} \intt \qP \e^{-\ii \qp_\qmu \qx^\qmu} \qpsip (\qp)
\f {\qm \bit \dd^3 \qqp}{\qp_0} =
\f{1}{(2\pi)^2} \intt \qP \e^{-\ii \qp_0 \qx_0 } \e^{ \ii
\qp_\qj \qx_\qj } \bit \qpsip (\qp) \f {\qm \bit \dd^3 \qqp}{\qp_0}
= \f{1}{\sqrt{2\pi}} \bit \qpsix(t, \qqx)
 }
\m {\9 2 { t \equiv x^0 }}. The momentum space condition \m { \qpsip \1
1 { \qp } \in \qNpp }, which can also be expressed in the two other
forms
 \eq{.5.38.}{
\qgam^\qmu \qp_\qmu \qpsip \1 1 { \qp } = \qm \qpsip \1 1 { \qp } ,
 \qquad
 \9 1 { \qalp_\qj \qp_\qj + \qbet \qm } \qpsip \1 1 { \qp } =
\qp_0 \qpsip \1 1 { \qp } ,
 }
is transformed to
 \eq{.5.39.}{
\qgam^\qmu (\ii\pd_\qmu) \qpsix \1 1 { \qx } = \qm \qpsix \1 1 { \qx } ,
 \qquad
\ii \pd_\qt \qpsix \1 1 { \qt, \qqx } = \9 2{ \qalp_\qj \9 1 { - \ii
\pd_\qj } + \qbet \qm } \qpsix \1 1 { \qt, \qqx } .
 }
Because of the unitary three-Fourier transform inside \re{.5.37.}, the
coordinate space integral
 \eq{.5.40.}{
\intt{\qX} \9 1 { \qpsix^\qdag \qpsix } \1 1 {t, \qqx} \bitt \dd^3 \qqx
 }
is time independent along the solutions of \re{.5.39.}, and corresponds
to the squared norm of \m { \qpsip } in the momentum space Hilbert space
\m { \qHfpp }. \1 2 {This time independence can also be seen from the
conserved probability current to be introduced in \re{.5.42.}.} The
coordinate space scalar product related to \re{.5.40.},
 \eq{.5.41.}{
\Scal{\qpsix_1 }{\qpsix_2 } = \intt{\qX} \qpsix_1^\qdag
\qpsix_2^{\vphantom{\qdag}} \bitt \dd^3 \qqx ,
 }
is also time independent along solutions. Accordingly, spatially
integrable solutions of \re{.5.39.}, endowed with \re{.5.41.}, form the
Hilbert space \m { \qHfxp } of the free spin-1/2 particle. \m { \qHfxp }
is the coordinate space equivalent of the momentum space Hilbert space
\m { \qHfpp }.

Physical quantities of the particle can be realized 
on \m { \qHfxp }. One interesting example is that of position
\cite{FarKurWei}.

For solutions of \re{.5.39.}, the four-current
 \eq{.5.42.}{
\qqj^\qmu = \qbar{\qpsix} \qgam^\qmu \qpsix 
 }
is conserved, that is, its four-divergence \m { \pd_\qmu \qqj^\qmu } is
zero. The timelike component,
 \eq{.5.43.}{
\qqj^0 = \qbar{\qpsix} \qgam^0 \qpsix = \qbar{\qpsix} \qbet \qpsix =
\qpsix^\qdag \qpsix ,
 }
is positive definite and is the probability density in the integral
\re{.5.40.}, i.e., \m { j^\mu } is the conserved probability current.

Actually, \re{.5.39.} is the Euler--Lagrange equation stemming from the
Lagrangian
 \eq{.5.44.}{
\qqL = \qbar \qpsix \9 2 { \ii \qgam^\qmu \pd_\qmu - \qm } \qpsix ,
 }
and \m { \qqj } is the conserved Noether current derivable from \m {
\qqL } corresponding to the global gauge transformations \m { \qpsix
\mapsto \e^{\ii \qchi} \qpsix }.

Solutions of \re{.5.39.}, i.e., of
 \eq{.5.45.}{
\9 1 { \ii \qgam^\qmu \pd_\qmu - \qm } \qpsix = 0 ,
 }
satisfy the free Klein--Gordon equation componentwise,
 \eq{.5.46.}{
\9 1 { \pd^\qmu \pd_\qmu - \qm^2 } \qpsix = 0 ,
 }
as follows by acting on \re{.5.45.} by \m { \ii \qgam^\qmu \pd_\qmu +
\qm } from the left -- as well as from \re{.4.25.}.

Should one start with \re{.5.46.}, selecting the \quot{positive}
solutions \1 2 {solutions of \re{.5.39.}} is not straightforward --
instead, it is the momentum space equivalent \re{.4.25.} that is
advantageous for this purpose.

Transforming \re{.3.23.} to coordinate space leads to
 \eq{.5.47.}{
\ii\bit \pd_0 \qpsix = \sqrt{- \triangle + m^2} \bittt \qpsix ,
 }
where the square root of the positive operator is well-defined.
Nevertheless, \re{.5.47.} is technically inconvenient, and is not
suitable for generalization of this free particle theory to nonfree
ones.

\subsection{Particle in external field}\label{.6..3.3.}

The generalization to nonfree cases, more specifically, to motion under
the action of a four-potential external field \m { \qA }, can be done
via minimal coupling in \re{.5.39.}, that is, via the substitution
 \eq{.6.48.}{
\pd_\qmu \bitttttt \qrepl \bitttttt \pd_\qmu + \ii \qe \qA_\qmu ,
 }
yielding
 \eq{.6.49.}{
\qgam^\qmu \9 1 { \ii\pd_\qmu - \qe \qA_\qmu } \qpsix = \qm \qpsix,
 \qquad
\ii \pd_0 \qpsix = \9 2{ \qalp_\qj \9 1 { - \ii \pd_\qj - \qe \qA_\qj }
+ \qbet \qm + \qe \qA_0 } \qpsix .
 }
The four-current \m{ \qqj^\qmu = \qbar{\qpsix} \qgam^\qmu \qpsix }
remains conserved under the more general equation \re{.6.49.}, and its
positive definite timelike component ensures a Hilbert space structure
for the solutions of \re{.6.49.} like in the free case. Similarly,
\re{.6.49.} is the Euler--Lagrange equation corresponding to the
Lagrangian
 \eq{.6.50.}{
\qqL = \qbar \qpsix \9 2 { \qgam^\qmu \9 1 { \ii \pd_\qmu - \qe \qA_\qmu
} - \qm } \qpsix ,
 }
and \m { \qqj } is still the conserved Noether current corresponding to
global gauge transformations.

On the other side, the minimally coupled Klein--Gordon equation is not
satisfied,
 \eq{.6.51.}{
\9 2 {  \9 1 { \pd^\qmu + \ii \qe \qA^\qmu } \9 1 { \pd_\qmu + \ii 
\qe \qA_\qmu } - \qm^2 } \qpsix \ne 0 .
 }
In fact, acting on \m { \9 2 { \qgam^\qnu \9 1 { \ii\pd_\qnu - \qe
\qA_\qnu } - \qm } \qpsix = 0 } by \m { \9 2 { \qgam^\qmu \9 1 {
\ii\pd_\qmu - \qe \qA_\qmu } + \qm } } gives a second order equation
that differs from the minimally coupled Klein--Gordon one by
 \eq{@52.}{
\2 2 { \qgam^\qmu \9 1 { \ii\pd_\qmu - \qe \qA_\qmu } + \qm }
\2 2 { \qgam^\qnu \9 1 { \ii\pd_\qnu - \qe \qA_\qnu } - \qm } \qpsix
&
\non\\ \label{.6.52.}
\mathrel{-}
\qg^{\qmu \qnu} \9 2 { \9 1 { \pd_\qmu + \ii \qe \qA_\qmu } \9 1 {
\pd_\qnu + \ii \qe \qA_\qnu } - \qm^2 } \qpsix
& \\ \non
& = \ii \qe \9 1 { \qg^{\qmu \qnu} \qqone - \qgam^\qmu \qgam^\qnu }
\9 1 { \pd_\qmu \qA_\qnu } \qpsix \ne 0 .
 }

\section{Spin 0 quantum mechanics}\label{.6..4.}


The antisymmetric spinor tensor formulation of the spin 0 case can be
done in close analogy to the spin-1/2 situation. The analogy is actually
so strong that it is enough to compare the two cases in form of a table
that lists the steps in brief form.

\subsection{Free particle, momentum space}\label{.7..4.1.}

\nc\qheaderrow{\qheader{SPIN 1/2}{SPIN 0\vphantom{/}}}

\y3 ex
\tabcolsep 1em  
 \begin{tabular}{ll}
 \qheaderrow
 \qrow{
\m { \qC^4 } valued \m { \qpsip }
 }{
\m { \qC^4 \wedge \qC^4 } valued \m { \qzetp }
 \\
\null\hfill
\1 1 {antisymmetric \m { 4 \times 4 } complex matrix valued}
 }\qrow{
\m{ \qgam^\qmu \qp_\qmu \qpsip = \qm \qpsip }
 }{
\m{ \qgam^\qmu \qp_\qmu \qzetp = \qm \qzetp }
 }\qrow{
\m { \qpsip \1 1 { \qp } \in \qNpp }
 }{
\m { \qzetp \1 1 { \qp } \in \qNpp \wedge \qNpp }
 }\qrow{
\m{\rc\arraystretch{1 } \qpsip \2 1 { \qpn } = \qmat{ \qpsip_1 \2 1 {
\qpn } \qnl \qpsip_2 \2 1 { \qpn }  \qnl  0  \qnl  0 }
 }
 }{
\m{\rc\arraystretch{1 } \qzetp \2 1 { \qpn } = \qmat{ 0 & \qzetp_{12} \2
1 { \qpn } & 0 & 0 \\ - \qzetp_{12} \2 1 { \qpn } & 0 & 0 & 0 \\ 0 & 0 &
0 & 0 \\ 0 & 0 & 0 & 0 } }
 }\qrow{
2 degrees of freedom:
 \\
\m{ \9 1 { \qgam^\qmu \qp_\qmu - \qm } \9 1 { \qc_1 \qn_1 +
\cdots + \qc_4 \qn_4 } = 0 }
 \\
\m{\Longrightarrow}\; only \m{\qc_1}, \m{\qc_2} can be nonzero
 }{
\m { 2 \wedge 2 = 1 } degree of freedom:
 \\
\m { \1 1 { \qgam^\qmu \qp_\qmu - \qm } ( \qc_{12} \bit \qn_1
\biT\wedge\biT \qn_2 + \qc_{13} \bit \qn_1 \biT\wedge\biT \qn_3 +
\cdots }
 \\
\m { \mathrel{+} \qc_{34} \bit \qn_3 \bi\wedge\bi \qn_4 ) = 0 }
\;\;\m{\Longrightarrow}\;\:only \m { \qc_{12} } can be nonzero
 }\qrow{
\m{ \9 1 { \qU_{\1 1 { \qa, \qL }} \qpsip } \1 1 { \qp } =
\e^{ \ii \qp_\qmu \qa^\qmu } \qD_\qL \qpsip \1 1 { \qL^{-1} \qp } }
 }{
\m{ \9 1 { \qU_{\1 1 { \qa, \qL }} \qzetp } \1 1 { \qp } =
\e^{ \ii \qp_\qmu \qa^\qmu } \biT \9 1 { \qD_\qL \otimes \qD_\qL }
\qzetp \1 1 { \qL^{-1} \qp }
 \\
\phantom{\9 1 { \qU_{\1 1 { \qa, \qL }} \qzetp } \1 1 { \qp } } = \e^{
\ii \qp_\qmu \qa^\qmu } \qD_\qL^{} \qzetp \1 1 { \qp } \qD_\qL^\qtrans }
 }\qrow{
irreducible unitary ray representation of
 \\
the
Poincar\'e group \1 1 {free elementary object}
 }{
irreducible unitary ray representation of
 \\
the
Poincar\'e group \1 1 {free elementary object}
 }\qrow{
\m{\rc\arraystretch{1 }
\qpsip^{\qW}
 = \qW
 \qpsip
= \qmat{ \qpsip_1^{\qW} \qnl \qpsip_2^{\qW} \qnl 0 \qnl 0 }
 }
 }{
\m{\rc\arraystretch{1 }
\qzetp^{\qW} = \qW \bit \qzetp \bit \qW^\qtrans =
\qmat{ 0 & \qzetp_{12}^\qW & 0 & 0 \\ - \qzetp_{12}^\qW & 0 & 0 & 0 \\
0 & 0 & 0 & 0 \\ 0 & 0 & 0 & 0 }
 }
 }\qrow{
\m{ \qp_0 \qpsip = \sqrt{ \qp_\qj \qp_\qj + \qm^2 } \bittt \qpsip }
 }{
\m{ \qp_0 \qzetp = \sqrt{ \qp_\qj \qp_\qj + \qm^2 } \bittt \qzetp }
 }\qrow{
\m{ \9 1 { \qp^\qmu \qp_\qmu - \qm^2 } \qpsip = 0 }
 }{
\m{ \9 1 { \qp^\qmu \qp_\qmu - \qm^2 } \qzetp = 0 }
 }\qrow{
\m{ \9 1 { \qp^\qmu \qp_\qmu - \qm^2 } \qpsip^{\qW} = 0 }
 }{
\m{ \9 1 { \qp^\qmu \qp_\qmu - \qm^2 } \qzetp^{\qW} = 0 }
 }\qrow{
\m{ \mS0 \rule[-4ex]{0em}{0ex}
\intt {\qPp} \qbar{\qpsip} \qpsip \f {\qm \bit \dd^3 \qqp}{\qp_0} =
\intt {\qPp} {\qpsip}^\qdag \qpsip \f {\qm^2 \bi \dd^3 \qqp}{\qp_0^2} }
 \\
\m{ \mS0
\phantom{ \intt {\qPp} \qbar{\qpsip} \qpsip \f {\qm \bit \dd^3
\qqp}{\qp_0} } = \intt {\qPp}\9 1 { \9 6 { \qpsip_1^{\qW} }^2 +
\9 6 { \qpsip_2^{\qW}}^2} \f {\qm^2 \bi \dd^3 \qqp}{\qp_0^2}
}
 }{
\m{ \mS0 \rule[-4ex]{0em}{0ex}
\intt {\qPp} \f {1}{2} \Tr \0 1 { \qbar{\qzetp} \qzetp }
\f {\qm \bit \dd^3 \qqp}{\qp_0} = \intt {\qPp} \f {1}{2}
\Tr \0 1 { {\qzetp}^\qdag \qzetp } \f {\qm^3 \bi \dd^3 \qqp}{\qp_0^3} }
 \\
\m{ \mS0
\phantom{ \intt {\qPp} \f {1}{2} \Tr \0 1 { \qbar{\qzetp} \qzetp } \f
{\qm \bit \dd^3 \qqp}{\qp_0} } = \intt {\qPp} \9 6 {
\qzetp_{12}^{\qW}}^2 \f {\qm^3 \bi \dd^3 \qqp}{\qp_0^3} }
 }\qrow{
\m { \qHfpp }
 }{
\m { \qHffpp }
 }
 \end{tabular}

\subsection{Free particle, coordinate space}\label{.8..4.2.}

\y3 ex
 \begin{tabular}{ll}
 \qheaderrow
 \qrow{
\m { \qgam^\qmu (\ii\pd_\qmu) \qpsix = \qm \qpsix }
 }{
\m { \qgam^\qmu (\ii\pd_\qmu) \qzetx = \qm \qzetx }
 }\qrow{
\m { \ii \pd_\qt \qpsix = \9 2{ \qalp_\qj \9 1 { - \ii \pd_\qj } + \qbet
\qm } \qpsix }
 }{
\m { \ii \pd_\qt \qzetx = \9 2{ \qalp_\qj \9 1 { - \ii \pd_\qj } + \qbet
\qm } \qzetx }
 }\qrow{
\m { \qqL = \qbar \qpsix \9 1 { \ii \qgam^\qmu \pd_\qmu - \qm } \qpsix }
 }{
\m { \qqL = \f {1}{2} \Tr \9 2 { \bittt \qbar{\qzetx} \9 1 { \ii \qgam^\qmu
 \pd_\qmu - \qm } \qzetx } }
 }\qrow{
\m { \qqj^\qmu = \qbar{\qpsix} \qgam^\qmu \qpsix
}\quad conserved Noether current
 }{
\m { \qqj^\qmu = \f {1}{2} \bi \Tr \bi \9 1 { \bitt \qbar{\qzetx}
\qgam^\qmu \qzetx \bit }
}\hfill
conserved Noether current
 }\qrow{
\m { \qqj^0 = \qpsix^\qdag \qpsix = \9 6 { \qpsix_1^{\qW} }^2 + \9 6 {
\qpsix_2^{\qW} }^2 \ge 0
 }
 }{
\m { \qqj^0 = \f {1}{2} \biT \Tr \2 1 { \qzetp^\qdag \qzetp } =
\9 6 { \qzetx_{12}^{\qW}}^2 \ge 0
 }
 }\qrow{
\m { \Scal{\qpsix_1
}{\qpsix_2
} = \intt{\qX} \qpsix_1^\qdag \qpsix_2^{} \bittt \dd^3 \qqx }
 }{
\m{ \Scal{\qzetx_1
}{\qzetx_2
} = \intt{\qX} \f {1}{2} \Tr \2 1 { \qzetp_1^\qdag \qzetp_2^{} } \bitt
\dd^3 \qqx }
 }\qrow{
\m { \qHfxp }
 }{
\m { \qHffxp }
 }\qrow{
\m { \9 1 { \pd^\qmu \pd_\qmu - \qm^2 } \qpsix = 0 }
 }{
\m { \9 1 { \pd^\qmu \pd_\qmu - \qm^2 } \qzetx = 0 }
 }\qrow{
\m { \9 1 { \pd^\qmu \pd_\qmu - \qm^2 } \qpsix^{\qW} = 0 }
 }{
\m { \9 1 { \pd^\qmu \pd_\qmu - \qm^2 } \qzetx^{\qW} = 0 }
 }\qrow{
\m { \ii\bit \pd_0 \qpsix = \sqrt{- \triangle + m^2} \bittt \qpsix }
 }{
\m { \ii\bit \pd_0 \qzetx = \sqrt{- \triangle + m^2} \bittt \qzetx }
 }\qrow{ \baselineskip4.3 ex
\m { \ii\bit \pd_0 \qpsix^{\qW} = \sqrt{- \triangle + m^2} \bittt \qpsix^{\qW} }
 \\
\m { \ii\bit \pd_0 \qpsix_1^{\qW} = \sqrt{- \triangle + m^2} \bittt
\qpsix_1^{\qW} }
 \\
\m { \ii\bit \pd_0 \qpsix_2^{\qW} = \sqrt{- \triangle + m^2} \bittt
\qpsix_2^{\qW} }
 }{ \baselineskip4.3 ex
\m { \ii\bit \pd_0 \qzetx^{\qW} = \sqrt{- \triangle + m^2} \bittt \qzetx^{\qW} }
 \\
\m { \ii\bit \pd_0 \qzetx_{12}^{\qW} = \sqrt{- \triangle + m^2} \bittt
\qzetx_{12}^{\qW} }
 \\
equivalence to scalar Klein--Gordon theory
 \\
via \mm { \qzetx_{12}^{\qW} \equiv \qphix }
 }
 \end{tabular}

\subsection{Particle in external field}\label{.8..4.3.}

\y3 ex
 \begin{tabular}{ll}
 \qheaderrow
 \qrow{
\m { \qgam^\qmu \9 1 { \ii\pd_\qmu - \qe \qA_\qmu } \qpsix = \qm \qpsix }
 }{
\m { \qgam^\qmu \9 1 { \ii\pd_\qmu - \qe \qA_\qmu } \qzetx = \qm \qzetx }
 }\qrow{
\m { \ii \pd_0 \qpsix = \9 2{ \qalp_\qj \9 1 { \ff{1}{\ii} \pd_\qj - \qe
\qA_\qj } + \qbet \qm + \qe \qA_0 } \qpsix }
 }{
\m { \ii \pd_0 \qzetx = \9 2{ \qalp_\qj \9 1 { \ff{1}{\ii} \pd_\qj - \qe
\qA_\qj } + \qbet \qm + \qe \qA_0 } \qzetx }
 }\qrow{
\m { \qqL = \qbar \qpsix \9 2 { \qgam^\qmu \9 1 { \ii \pd_\qmu - \qe
\qA_\qmu } - \qm } \qpsix }
 }{
\m { \qqL = \f {1}{2} \Tr \9 3 { \bittt \qbar{\qzetx} \9 2 {
\qgam^\qmu \9 1 { \ii \pd_\qmu - \qe \qA_\qmu } - \qm } \qzetx } }
 }\qrow{
\m { \qqj^\qmu = \qbar{\qpsix} \qgam^\qmu \qpsix
}\quad conserved Noether current
 }{
\m { \qqj^\qmu = \f {1}{2} \bi \Tr \bi \9 1 { \bitt \qbar{\qzetx}
\qgam^\qmu \qzetx \bit }
}\hfill
conserved Noether current
 }\qrow{
\m { \qqj^0 = \qpsix^\qdag \qpsix \ge 0
 }
 }{
\m { \qqj^0 = \f {1}{2} \biT \Tr \2 1 { \qzetp^\qdag \qzetp } \ge 0
 }
 }\qrow{
\m { \Scal{\qpsix_1
}{\qpsix_2
} = \intt{\qX} \qpsix_1^\qdag \qpsix_2^{} \bittt \dd^3 \qqx }
 }{
\m{ \Scal{\qzetx_1
}{\qzetx_2
} = \intt{\qX} \f {1}{2} \Tr \2 1 { \qzetp_1^\qdag \qzetp_2^{} } \bitt
\dd^3 \qqx }
 }\qrow{ \baselineskip3.9 ex
\m { \9 2 { \9 1 { \pd^\qmu + \ii \qe \qA^\qmu } \9 1 { \pd_\qmu + \ii
\qe \qA_\qmu } - \qm^2 } \qpsix \ne 0 }
 \\
\null\hfill\1 2 {see \re{.6.52.}}
 }{ \baselineskip3.9 ex
\m { \9 2 { \9 1 { \pd^\qmu + \ii \qe \qA^\qmu } \9 1 { \pd_\qmu + \ii
\qe \qA_\qmu } - \qm^2 } \qzetx \ne 0 }
 \\
\null\hfill\1 2 {see \re{.6.52.}, with replacement \mm { \qpsix \qrepl
\qzetx } }
 }
 \end{tabular}

\section{Outlook -- a list of tasks for the future}\label{.9..5.}


Having established the relationship between the spinor tensor formalism
and the Klein--Gordon scalar one in the free case, some further
investigation between the two could give some further insight. For
example, the Klein--Gordon conserved current could be expressed in terms
of the spinor tensor formalism, while the spinor tensor conserved current
could be transformed to the Klein--Gordon scalar language.

The relationship between the known Klein--Gordon scalar version of the
Foldy--Wouthuysen transformation \cite{Case}
\1 1 {formulated in terms of the Feshbach--Villars form}
and the one shown here \1 1 {adaptation of the spin-1/2 version} would
be interesting to investigate.

One of the most exciting questions is the spectrum of the Coulomb problem
in the spinor tensor theory.

Second quantization of the free case would be important and seems
straightforward.

Concerning interacting quantum field theories in general, we expect some
new interaction and self-interaction terms \1 1 {recall that the
Foldy--Wouthuysen relationship between the spinor tensor and the
Klein--Gordon scalar is \quot{nonlocal} in the free case, and even this
type of equivalence is broken for nonfree situations}. Renormalization
properties of the spinor tensor field might be better than what
superficial power counting says because of possible cancellations due to
antisymmetricity of the field.

As one concrete example, the spinor tensor QED may behave differently
from the Klein--Gordon based \quot{scalar QED}.

Another interesting idea is to realize interaction terms through which
the Higgs particle can give mass to fermions, and to explore Higgs
self-interaction potentials.

As another direction of outlook, similarly to the version of zero-spin
quantum mechanics presented here, the treatment of higher spin particles
is also worth revisiting. For example, the spin-1 particle is known to
admit a symmetric spinor tensor description \1 1 {see a historical
overview in \cite{Greiner}}. Issues like \quot{the wave function of the
photon} and \quot{position operator of the photon} as well as possible
quantum field theoretical benefits motivate such a line of
investigation.

\subsection*{Acknowledgement}

Support NKFIH K116375 is acknowledged.

\end{document}